\documentclass[11pt,a4paper]{article}
\pdfoutput=1

\usepackage{jheppub}
\usepackage{epsfig}
\usepackage{latexsym}
\usepackage{amsfonts}
\usepackage{amsmath}
\usepackage{amsthm}
\usepackage{amssymb}
\usepackage{amsbsy}
\usepackage{multirow}
\usepackage{braket}
\usepackage{tikz}
\usepackage{enumerate}
\usetikzlibrary{matrix,arrows}

\newcommand{\tr}{\mbox{tr}}

\newcommand{\CC}{\mathbb{C}} 
\newcommand{\ZZ}{\mathbb{Z}} 

\def\calo         {{\cal O}}

\def\calw         {{\cal W}}
\newsavebox{\uuunit}
\sbox{\uuunit}
    {\setlength{\unitlength}{0.825em}
     \begin{picture}(0.6,0.7)
        \thinlines
        \put(0,0){\line(1,0){0.5}}
        \put(0.15,0){\line(0,1){0.7}}
        \put(0.35,0){\line(0,1){0.8}}
       \multiput(0.3,0.8)(-0.04,-0.02){12}{\rule{0.5pt}{0.5pt}}
     \end {picture}}

\def\be{\begin{equation}}
\def\ee{\end{equation}}
\def\bea{\begin{eqnarray}}
\def\eea{\end{eqnarray}}

\def\a{\alpha}

\def\g{\gamma}

\def\d{\delta}

\def\l{\lambda}
\def\L{\Lambda}
\def\th{\theta}

\def\f{\phi}

\def\p{\pi}
\def\r{\rho}

\def\t{\tau}

\def\sF{{{ F}\!\!\!\!\hskip.8pt\hbox{\raise1pt\hbox{/}}\,}}
\def\som{{{ \omega}\!\!\!\!\hskip.8pt\hbox{\raise1pt\hbox{/}}\,}}
\def\sJ{{{\rm J}\!\!\!\!\hskip.8pt\hbox{\raise1pt\hbox{/}}\,}}

\def\to{\rightarrow}
\def\nonu{\nonumber \\{}}
\def\half{{1 \over 2}}

\def\gll{gl [ \l ]}
\def\hsl{hs [ \l ]}
\def\cwl{\calw_\infty [\l ]}

\def\lt{{\lambda_{H}}}

\title{A note on conical solutions in 3D Vasiliev theory}
\author[1]{Andrea Campoleoni,}
\author[2]{Tom\'{a}\v{s} Proch\'{a}zka}
\author[2]{and Joris Raeymaekers}
\affiliation[1]{Universit\'e Libre de Bruxelles and International Solvay Institutes, ULB-Campus Plaine CP231, B-1050 Brussels, Belgium.}
\affiliation[2]{Institute of Physics of the ASCR,
Na Slovance 2, 182 21 Prague 8, Czech Republic.}
\emailAdd{andrea.campoleoni@ulb.ac.be}
\emailAdd{prochazkat@fzu.cz}
\emailAdd{joris@fzu.cz}

\abstract{
We construct a class of smooth solutions in three-dimensional Vasiliev higher spin theories based on the gauge algebra $hs[\lambda]$.
These solutions naturally generalize the previously constructed conical defect solutions in higher spin  theories with $sl(N)$ gauge algebra, to which they reduce when
$\lambda$ is taken to be equal to $N$.
We provide evidence for their identification  with specific primary states
 of the $\calw_\infty [\lambda]$ algebra in a particular classical limit. In terms of the
 Gaberdiel-Gopakumar-'t Hooft limit of the $W_N$ minimal models, this limit corresponds to a regime where the 't Hooft coupling becomes large.}

\begin{document}
\maketitle

\section{Introduction}

In this paper we focus on the gauge sector of the three-dimensional higher-spin gauge theories coupled to matter constructed by Prokushkin and Vasiliev \cite{Prokushkin:1998bq,Prokushkin:1998vn}.
The simplest bosonic example of these systems describes a tower of massless fields of spin $2,3, \ldots$  interacting with two (generically massive) spin 0 fields. This theory was proposed in \cite{Gaberdiel:2010pz} to be holographically dual to the 't Hooft limit of the $W_N$ minimal models  (see \cite{Gaberdiel:2012uj} for a review of the subsequent refinements of the original proposal and a more complete list of references). Various generalizations of this basic setup can be considered, such as a supersymmetrization including two (generically massive) spin 1/2 fields and a tower of massless fermions of spin $3/2, 5/2, \ldots$, incorporating a spin 1 Chern-Simons field, and the inclusion of internal Chan-Paton like degrees of freedom \cite{Prokushkin:1998bq}. Some of these extensions have been proposed to be holographically dual to specific 2D conformal field theories \cite{Creutzig:2011fe,Candu:2012jq}.

These theories are naturally formulated in Anti-de Sitter  space and depend on 2 dimensionless parameters.
One of these is the Brown-Henneaux central charge $c$ proportional the ratio of the Anti-de Sitter length scale $l$ and the 3D Newton constant $G$ \cite{Henneaux:2010xg,Campoleoni:2010zq,Gaberdiel:2011wb}. We will
 assume $c$ to be large in order to be in the weakly interacting classical regime. The second dimensionless parameter, $\l$, arises as the vev of one of the auxiliary fields and sets the mass of the spin 0 and spin 1/2 fields in AdS units \cite{Prokushkin:1998bq}.
The parameter $\l$ also determines the bosonic part of the algebra of higher spin gauge symmetries to be the infinite-dimensional Lie algebra $\hsl$,\footnote{We will not consider here the cases in which the higher spin algebra is a subalgebra of $\hsl$, such as the even spin projections considered in \cite{Ahn:2011pv,Gaberdiel:2011nt}.} which can be seen as a continuation  of $sl(N)$ where $N \to  \l$ \cite{Feigin} as we will review below. The Prokushkin-Vasiliev field equations admit a consistent truncation to the sector where only the massless fields of spin 2 and higher are turned on \cite{Ammon:2011ua}. This subsector is governed by a Chern-Simons theory with $hs[\l]$ gauge symmetry \cite{Blencowe:1988gj,Bergshoeff:1989ns}. When $\l = N$ the Killing form of $\hsl$ degenerates and the Chern-Simons action reduces to that of a $sl(N)$ theory.

In this note we discuss a class of smooth solutions of the $\hsl$ Chern-Simons field equations that naturally generalize to arbitrary values of $\l$ the conical solutions  of the $sl(N)$ Chern-Simons theory proposed in \cite{Castro:2011iw} (for subsequent work on conical solutions see \cite{Tan:2012xi,Datta:2012km,Hikida:2012eu,Chen:2013oxa}). To this end we will first elaborate upon an explicit matrix description of $hs[\lambda]$ that allows one to accomodate our solutions and then we will present them. We eventually provide evidence for their identification  with specific primary states
 of the $\calw_\infty [\l]$ algebra in a particular classical limit.

\section{Chern-Simons subsector of Vasiliev theory}\label{sec:cs}

The consistent truncation of the Vasiliev theories that describes the sector where only massless fields of spin 2 and higher are turned on is governed by a Chern-Simons  theory with $\hsl$ gauge symmetry. We will here consider AdS higher spin gauge theories in Euclidean signature, which means that we will take as the gauge algebra a single complex $\hsl$  algebra rather than a direct sum of two copies of one of the real forms of $\hsl$ \cite{Witten:1989ip}. The reason for this choice is that, as explained in \cite{Castro:2011iw}, it leads to a richer spectrum of classical solutions which matches the dual CFT spectrum more closely.  The equations of motion simply  state that the connection is flat:
\be
F =d A + A \wedge  A =0 \, .\label{flatness}
\ee

We will now review some properties of the infinite-dimensional algebra $\hsl$, see \cite{Pope:1989sr} or the recent \cite{Gaberdiel:2011wb,Campoleoni:2011hg} for a more complete discussion.
 We label the $hs[\lambda ]$ generators as $T^\ell_m$ with integer spin ($\ell \geq 1$) and mode ($|m|\leq \ell$) indices.\footnote{We thus follow the conventions of \cite{Pope:1989sr,Campoleoni:2011hg}, while our generators can be mapped in the $V^s_m$ of \cite{Gaberdiel:2011wb} by $V^s_m = T^{s-1}_m$.} In order to deal with the Lie algebra $\hsl$ it is useful to add an ``identity'' element $T^0_0$ and consider the enlarged vector space $\gll = \CC \oplus \hsl$, where $\CC$ corresponds to the component along $T^0_0$. The resulting vector space can be identified with a suitable quotient of the enveloping algebra of $sl(2)$ and therefore it is naturally an associative algebra. The product of two generators can be described as \cite{Pope:1989sr}
\be \label{star-product}
T^{\,i}_m \star T^{\,j}_n \equiv \half \sum_{k\,=\,|i-j|}^{i+j} f_\l \! \left(
\begin{array}{ccc|cc}
i & j &&& k \\
m & n &&& m+n
\end{array}
\right)  T^{k}_{m+n} \, ,
\ee
where the $f_\l$ are the structure constants given explicitly in Appendix \ref{apphsl}. The element  $T^0_0$ plays the role of $\star$-identity. The  Lie bracket is simply the $\star$-commutator
\be
\left[\, T^{\,i}_m \,,T^{\,j}_n \,\right] \equiv\, T^{\,i}_m \star T^{\,j}_n \,-\, T^{\,j}_n \star T^{\,i}_m \, .
\ee

From the above considerations it follows that the Chern-Simons equations are formally invariant under finite gauge transformations of the form
\be
A \to g^{-1}\star (A+d)\star g \, ,
\ee
where $g= e_\star^v$ is the $\star$-exponential of an element of $\hsl$. We will in this work skirt around thorny questions such as for which linear combinations of the generators the  $\star$-exponential makes sense and if the resulting elements form a Lie group. We would like to point out however that the $\star$-exponential is
  well-defined on a multiple of a projector $P$ of the  $\star$-algebra satisfying $P\star P = P$, giving the result
\be
e_\star^{c P} = T^0_0 + ( e^{ c} - 1) P.\label{expproj}
\ee
This observation will play an important role in our construction of smooth solutions in Section \ref{condefs}.\footnote{For a review on the use of projectors in the construction of solutions of higher-spin field equations see e.g.~\cite{Iazeolla:2012nf}.}

Another important operation in the characterization of $\gll$ is the trace. For an element $v\in \gll$ it is defined to be proportional to the coefficient of $T^0_0$:
\be
\tr\, v = {6 \over  (\l^2-1)}\, v\,|_{T^\ell_m =\, 0\   {\rm for} \ \ell\,>\,0} \label{tr} \, .
\ee
The normalisation is chosen such that $\tr (T^1_1\, T^1_{-1}) = -1$. The definition of the trace gives a natural definition of the determinant of a $\star$-exponential:
\be
\det e_\star^v \equiv e^{{\rm tr} \, v} \, .
\ee
With this definition, if $v$ belongs to $\hsl$, the corresponding finite gauge transformation $g= e_\star^v$ has unit determinant.

The trace also induces an $\hsl$-invariant inner product $(u, v) = \tr (u \star v)$ \cite{Vasiliev:1989re}, that allows to define a Chern-Simons action and for which the chosen basis $\{ T^\ell_m \}$ is orthogonal. For generic $\l$ this inner product is nondegenerate, but when $\l$ is equal to an integer $N$ with $N \neq -1,0,1$\footnote{The normalization convention for the trace in (\ref{tr}) was in fact chosen such that the inner product is nondegenerate for $\l = -1,0,1$.}  it degenerates:
\be
\tr (T^{\,i}_m \star T^{\,j}_n ) = 0\qquad {\rm for}\ i,j \geq N \, .
\ee
This implies that an ideal appears, spanned by the generators $T^\ell_m$ for $\ell\geq N$. Quotienting by this ideal truncates to the algebra $sl(N)$.

It will also be useful to have at our disposal a concrete representation of $\hsl$   in terms of infinite matrices.
A representation of  $\hsl$ (analogous to the defining representation of $sl(N)$) can be constructed from a representation of $sl(2)$ with commutation relations
\bea
\left[\,J_+\,,J_-\,\right] &=& 2 J_0 \, ,\\
\left[\,J_0\,, J_{\pm}\,\right] &=& \mp J_\pm \, ,\label{sl2}
\eea
for which the quadratic Casimir takes the value
\be
C_2 = J_0^2 - \half (J_+ J_- + J_-J_+ )= {1 \over 4 } ( \l^2 -1) \, .\label{Cas}
\ee
For noninteger $\l$, such a representation is necessarily infinite-dimensional.
The basis elements $T^\ell_m$ of  $\gll$ are represented as elements of the enveloping algebra of this $sl(2)$ (following the conventions of \cite{Pope:1989sr}):
\be
T^\ell_m = (-1)^{\ell-m}\,
\frac{(\ell+m)!}{(2\ell)!} \,
 \Bigl[ \underbrace{J_-, \dots [\,J_-, [\,J_-}_{\hbox{\footnotesize{$\ell-m$ terms}}}, (J_+)^{\ell}\,]]\Bigr] \ .\label{univ}
\ee
Note that $T^0_0$ is the identity. The algebra  $\hsl$  is the subalgebra spanned by the generators $T^\ell_m$ with $\ell \geq 1$, and can be seen as a continuation  of the special linear algebra $sl(N)$ to arbitrary  $N$.

As recalled e.g.\ in \cite{Gaberdiel:2011wb}, for each non-integer $\l$ there are two highest weight representations of $sl(2)$ with highest weight $\frac{1}{2} (\pm \l - 1)$. The corresponding conjugate representations are lowest weight representations with lowest weight $\frac{1}{2} (\mp \l + 1)$. Each of these four representations can be used to build a faithful representation of $\hsl$ through \eqref{univ}. Concretely, we will use the following infinite matrices as  representatives for the $sl(2)$ elements \cite{Khesin:1994ey}:
 \bea
(J_+)_{jk} &=& \d_{j,\,k+1} \, , \\
(J_-)_{jk} &=& j (j-\l)\, \d_{j+1,\,k} \, , \\
(J_0)_{jk} &=&  \half (\l + 1 - 2j )\,\d_{j,\,k} \, , \label{J_0}
 \eea
 where the range of the indices is $j,k = 1,2,\ldots$.
One easily checks that these satisfy \eqref{sl2} and \eqref{Cas}, and that there is a highest weight vector of weight
 $\half (\l-1)$. From (\ref{univ}) one obtains an explicit representation for the $\hsl$ generators:
\be \label{gen_rep}
(T^\ell_m)_{jk} \,=\, (-1)^{\ell-m} \sum_{n\,=\,0}^{\ell-m}{\ell-m \choose n} \frac{\left[\,\ell\,\right]_n}{\left[\,2\ell\,\right]_n} \left[\,\ell-\l\,\right]_n \left[\,j-m-1\,\right]_{\ell-m-n}\, \d_{j,\,k+m} \ ,
\ee
where $[a]_k$ denotes the descending Pochhammer symbol
\be
[a]_k = a(a-1)\cdots(a-k+1) \, , \qquad \textrm{with}\ [a]_0\equiv1 \, .
\ee

From this expression, derived in Appendix \ref{apphsl}, one sees that the $T^\ell_m$, viewed as infinite matrices, have some special  properties.
The nonzero elements $(T^\ell_m)_{j,\,j-m}$ are $m$ spaces removed from the main diagonal, and furthermore they are polynomial in $j$.
When $\l$ is a natural number $N$, the $T^\ell_m$ have the block form
\be
T^\ell_m = \left( \begin{array}{cc} \t^\ell_m & 0 \\ * & * \end{array} \right)\label{ideal}
\ee
where the $\t^\ell_m$ are the generators of $gl(N)$ in the $N$-dimensional representation. This shows the appearance of the ideal mentioned earlier which is spanned by the generators $T^\ell_m$ for $\ell \geq N$.
Quotienting by it corresponds to projecting onto the $gl(N)$ generators $\t^\ell_m$.

Similar nice properties hold for finite linear combinations of the generators $T^\ell_m$. However for our purposes
 (namely the Drinfeld-Sokolov reduction of $\hsl$) it will not suffice to consider only  finite linear combinations of the $T^\ell_m$ but we shall need to allow also some well behaved infinite sums (see e.g.\ \eqref{hwgauge} below) as well as matrices where the polynomial property is
 violated in a finite number of rows (see e.g.\ \eqref{ugauge}).
A characterization of the resulting space of matrices was given  in \cite{Khesin:1994ey} which we will adhere to in this work.
We will consider a matrix $v$ to belong to $\gll$ if it satisfies the following properties:
\begin{enumerate}[(a)]
\item There exists a number $N$ such that $v_{j,\,k}=0$ if $j>k + N$.\label{ca}
\item The matrix elements along a diagonal, $v_{j,\,j+n}$ for some fixed $n$, become polynomial in $j$ for sufficiently large $j$.\label{cb}
\end{enumerate}

The trace (\ref{tr})  of  an element of $\gll$ can also be computed in the matrix representation  as follows \cite{Khesin:1994ey}:
\be
\tr \, v = {6 \over \l (\l^2 -1 )} \lim_{N \to \lambda} \sum_{j\,=\,1}^N v_{jj} \, , \label{trform}
\ee
with the proviso that $N$ has to be taken large enough that the resulting polynomial in $N$ stabilizes. The property (\ref{cb}) ensures that such $N$ always exists so that \eqref{trform} is an unambiguous definition. The trace \eqref{trform} is normalized such that it agrees with the alternative definition \eqref{tr} on products of the generators $T^\ell_m$. Moreover, the $\star$-product \eqref{star-product} agrees with the matrix product, so that in the following we will often omit the $\star$ symbol.

\section{Boundary and smoothness conditions}\label{secbc}
The $hs[\l ]$ Chern-Simons theory admits a vacuum solution describing global AdS$_3$ without higher spin fields. AdS$_3$ is topologically a solid cylinder, on which we choose a radial coordinate $\r$ such that the boundary is at $\r \to \infty$ and local complex coordinates $z, \bar z$ which parameterize a cylinder at surfaces of constant $\r$. We will interpret the complex coordinate as $z\equiv \f + i t_E$ where $t_E$ is the Euclidean time and $\f$ an angular coordinate with periodicity $\f \sim \f + 2 \p$.
More generally, we are interested in smooth connections defined on the solid cylinder which approach the global AdS solution near the boundary.  As explained in \cite{Henneaux:2010xg,Campoleoni:2010zq,Gaberdiel:2011wb,Campoleoni:2011hg}, after imposing AdS  boundary conditions and fixing the so-called ``highest weight'' gauge, the allowed flat connections are of the form
\bea
A &=& g^{-1} a(z) g\, dz + g^{-1} dg \, , \qquad g = e^{\r\, T^1_0} \, ,\label{rhodepgauge} \\[5pt]
a(z) &=& \a\,  T^1_1 + {12 \p \over c} \sum_{\ell = 1}^\infty {\a^{-\ell}\over N_\ell}\, W_{\ell+1} (z)\, T^{\ell}_{-\ell} \, ,\label{hwgauge}
\eea
where $\a$ is an arbitrary real constant which can be  absorbed in a shift of $\r$, $c = {3 l \over 2 G}$ is the Brown-Henneaux central charge, and the normalization constants are
  chosen to be $N_\ell = \tr\, T^\ell_{\ell}\, T^\ell_{-\ell}$.\footnote{The explicit expression is $$
N_\ell =   {3 \cdot 4^{-\ell}\sqrt{\pi}\,\Gamma(\ell+1)\over (\l^2-1)
\Gamma(\ell+\frac{3}{2})} (1-\l)_{\ell} (1+\l)_{\ell} \, ,
$$
where $(x)_n = \Gamma(x+n)/\Gamma(x)$ is the ascending Pochhammer symbol. In particular, we have $N_1 = -1$.} The allowed flat connections are characterized by holomorphic higher spin currents $W_s (z) $, which can be expanded
 in Fourier modes:
 \be
 W_s (z)= {1\over 2 \p} \sum \left( W^s_n - {c\over 24} \d_{s,2} \d_{n,0}\right) e^{- i n z}. \label{modeexp}
 \ee
 It was shown in \cite{Gaberdiel:2011wb,Campoleoni:2011hg} that under Poisson brackets the modes $W^s_n$  generate the classical $\calw_\infty^{\rm cl} [\l ]$  algebra\footnote{The procedure of constructing the Poisson brackets on the reduced phase space of asymptotically AdS connections is mathematically equivalent to the classical
 Drinfeld-Sokolov reduction \cite{Drinfeld:1984qv} of $hs[ \l ]$.}
 with central charge $c$. We also remark that the connection with all
 $W^s_n=0$ describes the global AdS vacuum.

 We are interested in solutions preserving time-translation and rotational invariance, and will therefore restrict to gauge fields where only the zero modes
 $W^s_0$ are allowed to be different from zero, in other words where  $a$ is a constant element of $hs[\l ]$. In what follows we will also  present solutions written down in a different gauge than the highest weight gauge (\ref{hwgauge}).
To read off charges in more general gauges, we need invariant expressions. For example, one can check  from (\ref{hwgauge}) that, for constant $a$, the energy and spin 3 charge are given by
\bea
h &\equiv& W^2_0 = {c \over 12} \left( \tr (a^2) + \half \right) , \label{weighttr}\\
W^3_0 &=& {c \over 18}\, \tr (a^3) \label{chargestr} \, .
\eea
The relation between higher spin charges and trace invariants can be extended recursively to arbitrary high spin using the identities (4.49) in \cite{Campoleoni:2011hg}. One useful property is that  sending $a\to - a$ is equivalent to changing the sign of all the odd charges $W^{2t+1}_0$. This is because (\ref{hwgauge}) implies the identity
\be
- a _{ \{W^{2t}_0,W^{2t+1}_0\} } = e^{i \p T^1_0} a_{ \{W^{2t}_0,- W^{2t+1}_0\} }  e^{-i \p T^1_0}.\label{mina}
\ee

Next we review the condition  imposed on the Chern-Simons  gauge field by requiring it to be smooth \cite{Castro:2011iw}.
Recall that in our parameterization the $\f$-circle is contractible,
 hence the requirement that the gauge field is smooth imposes that its holonomy  $H$ along the $\f$-cycle is a trivial  gauge transformation. Here, trivial means that it  acts trivially on all gauge fields, i.e. $H T^\ell_m H^{-1}= T^\ell_m$ for all $\ell,m$. Schur's lemma implies that   $H$ should be proportional to the unit element
\be
H=e^{2 \p a} = e^{i \varphi_0}\,  T_0^0, \label{smoothcond}
\ee
 for some (a priori complex) number $\varphi_0$.

\section{Conical solutions from $\star$-projectors}\label{condefs}
We now turn to the construction of smooth solutions in the $\hsl$  Chern-Simons theory. We will not construct our solutions in the highest weight gauge (\ref{hwgauge}) but rather in a  gauge where the connection $a$ is a linear  combination of the zero modes $T^\ell_0$ only. This is the most convenient gauge to compute $\star$-exponentials of $a$ and verify (\ref{smoothcond})  since, as we will see below, $a$ can be written as
a combination of commuting projection operators or, equivalently, $a$ is diagonal in the matrix representation introduced in Section \ref{sec:cs}.
 From the discussion in the previous sections, the sought-for diagonal gauge connections must satisfy the following requirements:
\begin{enumerate}
\item They are smooth, i.e.\ they satisfy (\ref{smoothcond}).\label{c1}
\item They can be brought to the highest weight gauge expression (\ref{hwgauge}) by a nonsingular gauge transformation.\label{c2}
\item When viewed as infinite matrices, they belong to the class discussed in Section \ref{sec:cs}. In the diagonal gauge this means that the condition (\ref{cb}) must be satisfied on the main diagonal.\label{c3}
\end{enumerate}
Of these conditions, \ref{c2}  is the most difficult to verify, since it involves checking invertibility of an infinite matrix. One can however derive necessary conditions for \ref{c2} to hold, which we do in Appendix \ref{gaugetransfo}. Therefore, in practice we will replace condition \ref{c2} with
\begin{enumerate}[1']\setcounter{enumi}{1}
 \item All eigenvalues of the diagonal connections are distinct and, for $\l$ non-integer, different from zero (see Appendix \ref{gaugetransfo}). Furthermore, \ref{c2} is satisfied  when $\l$ is taken to be an integer $N$ (for $N$ sufficiently large, in a sense to be discussed  below, in particular above (\ref{bsigma})).\label{c2p}
\end{enumerate}
We will now construct the most general diagonal gauge connection that satisfies the requirements \ref{c1}, \ref{c2p}'  and \ref{c3}. As we shall see, the simplest class of  such solutions is specified by a Young diagram, and these are the natural generalizations to $\hsl$ of the $sl(N)$ conical solutions constructed in \cite{Castro:2011iw}.

First we address the requirement \ref{c1}.  Instead of expanding in the standard basis $\{ T^\ell_0 \}$, it will be useful to expand in a different  basis for the diagonal gauge, namely the operators $\{ P_j \}$ which, in the matrix representation, are simply
 \be
(P_j)_{kl} = \d_{jk} \d_{jl}\, .\label{Ps}
\ee
The main advantage of the new  basis is that the $P_j$ are mutually commuting $\star$-algebra projectors:
\be
P_j P_k = P_j \d_{jk}\, .\label{commprojs}
\ee
This property is of course most easily verified in the matrix representation. Their trace, using (\ref{trform}) is\footnote{
Although the trace is distributive over a finite combination of $P_j$'s, we should warn the reader that this is not the case for infinite
sums of $P_j$'s. For example, one has $\tr \sum_{j=1}^\infty j^r P_j  = {6 \over \l (\l^2-1)} H^{(-r)}_\l$, where $H_n^{(r)}$ are the harmonic numbers.}
\be\tr\, P_j = {6 \over \l (\l^2-1)} \, . \label{trproj}\ee

The most general $\hsl$ element in the diagonal gauge can then be expanded as
\be
a = -\, i\, \sum_j m_j P_j + { \l^2 -1 \over 6}\, i\, \tr \left( \sum_j m_j P_j \right) T^0_0 \label{aproj}
\ee
where the second term is needed to ensure that $a$ is traceless and hence belongs to  $\hsl$.  Note that, from the relations $\sum_j P_j = T^0_0,\ \tr\, T^0_0 = 6/(\l^2-1)$, the coefficients $m_j$ are only determined up to an overall shift $m_j \to m_j + \a$ which leaves $a$ invariant. We can now exponentiate (\ref{aproj}) using (\ref{commprojs}) and (\ref{expproj}), to obtain
\be
e^{2 \p a} = e^{2\p{\l^2-1\over 6} i\, \tr (\sum_j m_j P_j)} \left( T^0_0 + \sum_j ( e^{-2 \p i m_j} - 1) P_j \right).
\ee
This is proportional to the identity $T^0_0$ if and only if, up to the overall shift freedom discussed above, the $m_j$ are integers.
This  conclusion could of course also be reached by using the matrix representation (\ref{Ps}).
We shall in what follows partially fix the shift freedom so that the $m_j$ are integers. This still leaves the freedom to shift all the $m_j$ by an
overall integer.
The connection then satisfies the trivial holonomy condition (\ref{smoothcond}) with $\varphi_0 = 2\p{\l^2 - 1 \over 6} \tr (\sum _j m_j P_j )  $.

Now we turn to the further conditions imposed by  requirements \ref{c2p}' and \ref{c3}. As discussed in Appendix \ref{gaugetransfo}, \ref{c2p}' imposes that
 all the $m_j$ are distinct and that, for $\l$ non-integer, the `eigenvalues' $m_j - {\l^2-1\over 6}  \tr (\sum_j m_j P_j)$ are different from zero. Condition \ref{c3} imposes that, for $j$ sufficiently large, the $m_j$ are given by a polynomial in $j$. The latter requirement implies that
 the $m_j$ are bounded, either from above or below. Indeed, because the  $m_j$ become polynomial for large $j$, depending on the sign of the coefficient of the highest power of $j$, there must exist an integer $M$ such
 that the $ m_j $  are either monotonically decreasing or increasing  for $j>M$. In the first case the $m_j$ are bounded above by $\sup { \{ m_j\}_{j\leq M}}$
 and in the second case they are bounded below by $\inf \{ m_j\}_{j\leq M}$.

 Let's first consider the case where the  integers $m_j$ are bounded from above.
By  performing a gauge transformation which permutes only the first $M$ eigenvalues, which can be achieved by embedding the appropriate $sl(M)$ Lie algebra element in $\hsl$,  we can arrange for the $m_j$ to form a strictly ordered set
\be
m_1> m_2 > \ldots
\ee
We then define new integers $s_j$
\be
s_j = m_j + j\label{ss}
\ee
which form an ordered set:
\be
s_1\geq s_2 \geq \ldots
\ee
Substituting (\ref{ss})  into (\ref{aproj}) and using the expression for $T^1_0 \equiv J_0$ in (\ref{J_0}) gives
\be
a = -\, i \sum_j s_j P_j + {\l^2 - 1 \over 6}\, i\,\tr \left(\sum_j s_j P_j \right)  T^0_0 - i\, T^1_0 \label{gensol} \, .
\ee

The condition \ref{c3} requires  that the $s_j$ become polynomial in $j$ for sufficiently large $j$.
The simplest class of solutions, which will have a natural interpretation in the dual CFT, is where this polynomial is simply a constant
integer $S$. The $s_j$  are then equivalent, upon performing an overall shift by $S$, to the positive natural numbers
\be
r_j = s_j - S.
\ee
Since the $r_j$ are decreasing and  become zero for sufficiently large $j$, they define a Young diagram $\L$  with a finite number of boxes,
containing $r_j$ boxes in the $j$-th row. In summary we have constructed a subclass of smooth solutions
in one-to-one correspondence with Young diagrams $\L$ given by
\be
a_\L = -\, i \sum_j r_j P_j + {i B \over \l}\, T^0_0 - i\, T^1_0 \label{asigma}
\ee
where $B$ denotes the total number of boxes in  $\L$ and we have used (\ref{trproj}).

We still have to verify the part of the condition \ref{c2p}' which imposes that, for $\l$ non-integer, none of the eigenvalues  are zero. For a given Young diagram this happens for a discrete set of values of $\l$, however  the range
$0 < \l < 1$ is special: for $\l$ in this range the eigenvalues are always nonzero.\footnote{Since the eigenvalues are $-i (r_j - j - B/\l + (\l+1)/2)$, a possible zero eigenvalue occurs at $\l\pm = \half( 2 (j-r_j) - 1 \pm \sqrt{
 (2 (j-r_j) - 1)^2 + 8B})$. The value  $\l_-$ is always negative, while $\l_+ \geq 1$. To show the latter inequality, we use that when $j-r_j>0$
 we have $\l_+ \geq 2(j-r_j) - 1 \geq1$, while for $j-r_j\leq0$   we use that from $B > r_j - j$ it follows that
 $8 B > 2 | 2 (j- r_j ) - 1|+1$.} This is also the range of $\l$  where the asymptotic symmetry algebra governing the Vasiliev theory is the same as the one governing the Gaberdiel-Gopakumar-'t Hooft limit of the $W_N$ minimal models.  We will discuss this in more detail in Section \ref{disc}.

Likewise, one can analyze the class of solutions where the $\{ m_j \}$ are bounded from below.
 In this case we can apply the steps of the previous paragraphs to the connection $-a$. Hence to every Young diagram
 $\L$ we can associate a second smooth solution, namely
 \be
 a_{ \overline{\L}} =  -\, a_\L =   i \sum_j r_j P_j - {i B \over \l}\, T^0_0 + i\, T^1_0 . \label{asigmabar}
 \ee
  From the remark above (\ref{mina}) we infer that
 $a_{ \overline{\L}}$ has the same even spin charges as $a_\L$ but that their odd spin charges have opposite signs.

Let's take a closer look at the sign of the energy of the solutions (\ref{asigma}), (\ref{asigmabar}).  The explicit expression for the energy is,
 from (\ref{weighttr}),
\be
h (a_\L) =  -   \left( B^2 - \l^2 B + \l \sum_j (c_j^2 - r_j^2 ) \right) {c \over 2 \l^2 (1- \l^2 )}\label{ensurpl}
\ee
where $r_j (c_j) $ denotes the number of boxes in the $j$-th row (column) and $B$ is the total number of boxes in $\L$.
Let's first consider the regime $|\l |>1$.  There are then solutions for which the energy is positive:  it
suffices to take $B> \l^2$ and $ \sum_j c_j^2 > \sum_j r_j^2$. This should be compared to the the $sl(N)$ case, where all the conical solutions have negative energy.
Once again the range $ |\l |\leq 1$ is special: the energy is always negative there, since the expression between brackets in (\ref{ensurpl}) is positive.
This can be seen by using the inequality $B^2 \geq \sum_j r_j^2 + 2 B$ for $0\leq \l \leq 1$ and the inequality $B^2 \geq \sum_j c_j^2 + 2 B$ for $-1\leq \l \leq 0$.

We conclude the discussion of the  solutions  (\ref{asigma}), (\ref{asigmabar})  with the important remark that they are natural continuations of the conical $sl(N)$ solutions of \cite{Castro:2011iw}. When $\l$ is taken to be a
positive integer $N$ larger than the number of rows in $\L$, projecting $a_\L$ and $a_{\overline{ \L}}$ to the first $N \times N$ block gives  the  $sl(N)$ matrices
\be
(b_{\L })_{jk} = -\,i \left(r_j - {B \over N} + {N + 1 \over 2}- j \right) \d_{jk};\ \ \ b_{ \overline{\L} } = - b_\L  \qquad j,k = 1,\ldots ,N \label{bsigma}
\ee
which are precisely the $sl(N)$ conical solutions constructed in \cite{Castro:2011iw}. Furthermore, (\ref{trform}) implies that all trace invariants
of $a_\L$ are continuations of those of $b_{\L }$
\be
\tr (a_\L )^n = \lim_{N\to \l} \tr_{N} (b_{\L })^n
\ee
 (and similarly for the trace invariants of $a_{\bar \L}$ and $b_{\bar \L}$), where $\tr_{N} $ is the $N \times N$ matrix trace in the normalization convention  $\tr_{N} b = {6 \over N (N^2-1)} \sum_{i=1}^N b_{ii}$. This property, implies that the classical $W$-charges $W^s_0$ of the $\hsl$ connections $a_\L$ are continuations of those of the $sl(N)$ connections $b_{\L }$ (provided that $N$ is larger than $B$), and likewise for the $a_{\overline{\L}}$.

The solutions (\ref{asigma}), (\ref{asigmabar}) constitute only the simplest class of solutions obeying \ref{c1}, \ref{c2p}', \ref{c3}.
We  now comment briefly on the more general solutions, which take the form (\ref{gensol}) where the $s_j$ become a non-constant polynomial in $j$
at large $j$. It will be interesting to see if these more general solutions also play a role in the higher spin/minimal model CFT correspondence.
 Let's give a simple example in this more general class, which was considered before in \cite{Kraus:2012uf}. Let's take
\be
s_j = -\,  n j \, .
\ee
for some integer $n$. The corresponding Chern-Simons connection is
\be
a_n = -\, i (n+1)\, T^1_0 \, .
\ee
For these solutions we can construct the explicit gauge transformation $g$ which brings them to the highest weight gauge (\ref{hwgauge}):
\be
a_n = {n+1 \over 2}\, g \left( T^1_1 + T^1_{-1} \right) g^{-1} \qquad {\rm with}\  g = e^{- {i\p\over 4} (T^1_1 - T^1_{-1})}.
\ee
From this and (\ref{weight}), (\ref{spin3}) we read off the energy and the higher spin charges:
\bea
W^2_0 &=& -\, {c \over 24}\, n(n+2) \, , \\
W^s_0 &=& 0 \qquad s>2 \, .
\eea
These solutions satisfy (\ref{smoothcond}) with $\varphi_0 =  \p (1- \l) (n+1) $. For example, for $\l = \half$ the central elements $e^{2 \p a_n}$  form the group $\ZZ_4$ \cite{Kraus:2012uf}.

\section{Identification with primaries of $\calw_\infty [\l]$}
As we recalled above, the asymptotic symmetry algebra of the Vasiliev theory is the classical algebra
$\calw_\infty^{cl} [ \l ]$ which arose from the classical Drinfeld-Sokolov reduction of $\hsl$. The quantum theory is  expected to be governed by the quantum $\calw_\infty [ \l ]$ algebra which arises as the quantum  Drinfeld-Sokolov reduction of $\hsl$.
This property lies at the core of the proposed AdS/CFT duality between 3D higher spin theories and 2D conformal theories with $\calw_\infty [ \l ]$
symmetry \cite{Gaberdiel:2010pz,Gaberdiel:2011wb,Gaberdiel:2012ku}. In this section we will propose an identification of the conical solutions constructed in the previous section with specific primaries of $\calw_\infty [ \l ]$.

As was explained in \cite{Gaberdiel:2012ku}, a class of interesting  degenerate  representations of $\calw_\infty [ \l ]$ can be obtained as continuations in $N$ of
the well-studied degenerate representations of the $W_N$ algebra. This is because $\calw_\infty [ N ]$ becomes equivalent to the $W_N$ algebra
after quotienting out a suitable ideal.\footnote{Note that $\calw_\infty [ \l ]$ also has representations which don't arise as continuations from $W_N$, because they are not compatible with the quotienting procedure for any $N$ \cite{Gaberdiel:2012ku}. We will not consider those here.} The degenerate representations of the $W_N$ minimal model at level $k$ are described by two Young diagrams with at most $N$ rows and at most $k$ and $k+1$ columns respectively.  For  any
two Young diagrams $\L^+$ and $\L^-$ with a finite number of boxes, there is a primary $(\L^+,\L^-)$ belonging to the  $W_N$ minimal model spectrum for sufficiently large values of $N$ and  $k$.
  Under analytic  continuation $N \to \l$ at fixed value of the central charge $c$,
 we obtain a primary of $\cwl$, which we will denote by  $(\L^+ ,\L^- )_\l$,
  whose higher spin charges are  continuations of those of the $W_N$ primaries.\footnote{This continuation is well-defined since the charges are polynomial in $N$ multiplied by powers of the quantities $\a_\pm$ defined in \eqref{thetaeq}, which admit a continuation in $\l$ as well.}
For example, the  energy and spin 3 charge can be expressed as (see e.g. \cite{Bouwknegt:1992wg})
\bea
h (\L^+,\L^-)_\l& = & \lim_{N \to \l}\,  \half \sum_{j = 1}^N\, (\th_j)^2    +  \frac{c-\l+1}{24} \label{weight}\\
W^3_0 (\L^+,\L^-)_\l & = & \lim_{N \to \l}\, {\g \over 3} \sum_{j = 1}^N\, (\th_j)^3 \label{spin3}
\eea
with the proviso that the upper limit $N$ of the sum  should be taken to be larger than the number of rows in $\L^+$ and $\L^-$.
The quantities appearing in these expressions are defined as follows.
We have introduced  an  infinite-dimensional vector $\theta$  constructed from the  Young diagram data:
\bea
\theta_j  &=& \a_+ \left( r^+_j - {B^+ \over \l} + {\l+1 \over 2} - j \right)   + \a_- \left( r^-_j - {B^- \over \l} + {\l+1 \over 2} - j \right) , \\
\a_+ &=& \sqrt{ \l + k + 1 \over \l + k} \, , \qquad \a_- = -\,\sqrt{ \l + k  \over \l + k+ 1} \, , \label{thetaeq}
\eea
where the `level' $k$ should be seen as a function of $\l$ and $c$  such that
 \be c = (\l-1) \left(1 - {\l(\l+1) \over (\l+ k)(\l+ k+1)} \right).\label{c}\ee
There are two roots of this equation, and in what follows we will choose the root\footnote{Choosing the other root, as was done e.g.\ in the conventions of \cite{Perlmutter:2012ds}, interchanges the role of $\L^+$ and $\L^-$.}
\be
 k(\l, c) = -\,\l - \half \left( 1 - \sqrt{ 1 + {4 \l(1-\l^2) \over c+1- \l }}\right).\label{kl}
\ee
Finally, $\g$ is a normalization constant ensuring that the spin 3 modes are  normalized in the same way as the classical spin 3 charges in (\ref{hwgauge}).\footnote{The explicit expression is $\g = -\sqrt{\frac{\l  \left(\l+ 2\right) \left(\l ^2-1\right)}{ \left((\text{c}-1) (\l +2)+\l ^2\right)}}$.}

Agreement with the classical Vasiliev theory is expected only upon taking the classical large $c$ limit, while keeping $\l$ fixed. For
the $\l=N$ case this is known as  the semiclassical limit \cite{Gaberdiel:2012ku}, and we will also call it that in the present context.  In this limit,
$\a_+$ and $\a_-$ are proportional to $c^{1/2}$ and $c^{-1/2}$ respectively, so that the contribution to the charges at leading order in $c$   depends only on $\L^+$.
The energy and spin 3 charge behave as
\bea
h (\L^+, \L^-)_\l &=&{ c \over 2 \l(1-\l^2)} \lim_{N \to \l}\,  \sum_{j = 1}^N \left(r^+_j - {B^+ \over \l} + {\l + 1 \over 2}- j \right)^2+ {c \over 24}+\calo(1)\nonu
&=& {c \over 12} \left( \tr (a_{\L^+} ^2) + \half \right) +\calo(1) \label{weightCFT}\\
W^3_0 (\L^+, \L^-)_\l &=& { i c \over 3 \l (1-\l^2)}  \lim_{N \to \l}\,  \sum_{j = 1}^N \left(r^+_j - {B^+ \over \l} + {\l + 1 \over 2}- j \right)^3+\calo(1) \nonu
&=& {c \over 18} \, \tr (a_{\L^+} ^3) +\calo(1)
\eea
where we have used (\ref{trform}), (\ref{asigma}). Comparing with (\ref{chargestr}),  we see that the order $c$ part of these charges agrees with the classical charges of the conical solution $a_{\L^+}$ of (\ref{asigma}), based on the Young tableau $\L^+$.
Furthermore, it is expected that for $\l = N$ the agreement between the  classical charges of the $sl(N)$ conical solutions and quantum higher spin charges of $W_N$ primaries established in \cite{Castro:2011iw} extends to all higher spin charges.
If this is the case, then the same agreement will hold for the classical charges of our $hs[ \l ] $ solutions and
  the higher spin charges of $\calw_\infty [ \l ]$  primaries because, as we have argued,  both sides are obtained by continuation in $N$.
  Further evidence for the CFT interpretation of the  $sl(N)$ conical solutions  comes from the matching of four-point functions \cite{Hijano:2013fja}, which seems likely to be amenable to the present $\hsl$ context.

 Note that from comparing only the charges, any primary $(\L^+ , \L^-)$ could correspond to the conical solution $a_{\L^+}$,  since $\L^-$
  does not affect the order $c$ part of the charges. However, for integer $\l$ it was argued in \cite{Perlmutter:2012ds} from comparing the classical symmetries
  of the  conical solutions with the behavior of null states at large $c$, that the correct choice is $\L^-=0$. In conclusion,  we have argued that the conical solution $a_{\L}$ constructed from the  Young tableau $\L$ is to be identified with the primary $(\L,0)_\l$ in the semiclassical limit.

So far we started from  $W_N$ primaries  obtained from Young diagrams  with a finite number of boxes, but we could also have considered Young diagrams with a finite number of antiboxes, i.e. diagrams $\overline{ \L}$  conjugate to diagrams $\L$ with a finite number of boxes.
The  primaries $(\overline{ \L},0 )_\l$  obtained in this way have the same even spin charges as $(\L ,0)_\l$ but have opposite odd spin charges.  This leads us to identify the $(\overline{ \L},0)_\l$ primaries with the smooth solutions
$a_{\bar \L}$ of (\ref{asigmabar}).

\section{Discussion}\label{disc}
We will now briefly discuss how the AdS/CFT dictionary between the Vasiliev theory   and $\cwl$ conformal field theory in the semiclassical limit
extends when matter is included, and comment on how this  relates to the 't Hooft limit of the $W_N$ CFT's in the duality proposal of \cite{Gaberdiel:2010pz}.

 Let's consider the Vasiliev theory in the regime $0\leq \l \leq1$.
 We have in this note constructed conical  solutions $a_{\L^+}$, and have argued that
 they are to be identified with the primaries $(\L^+,0)_\l$ of $\cwl$ in the semiclassical limit where $c$ is taken large at fixed
 $\l$. As we saw above, except for the global AdS solution corresponding to the vacuum  $(0,0)_\l$,  these primaries have negative energy in this  limit, and describe nonunitary representations of $\cwl$ at large $c$.
 The basic Vasiliev system also contains a  massive complex scalar which, with appropriate boundary conditions and  expanded around the
 AdS background, was argued \cite{Gaberdiel:2010pz} to correspond to the primary $(0, \square)_\l$ (and it's complex conjugate   to correspond to $(0, \overline{\square})_\l$). Multiparticle excitations of the scalar then build up the primaries of the form  $(0, \L^-)_\l$. Similarly, when expanded around a conical
 background  $a_{\L^+}$, an extension of the arguments in \cite{Perlmutter:2012ds} suggests that single- and multiparticle states of the scalar describe the
  primaries $(\L^+, \L^- )_\l$. Note that, in the regime $0\leq \l \leq1$, these states have positive energy above the conical background, see (\ref{scalperten}).  In this way the Vasiliev theory with AdS
boundary conditions appears to capture  all $(\L^+, \L^- )_\l$ primaries of $\cwl$, including those of the form $(\L, \L )_\l$, albeit in
 a nonunitary large $c$ limit.

Now let us discuss how this information relates to the Gaberdiel-Gopakumar-'t Hooft limit of the $W_N$ minimal models, which  governs the CFT of
the holographic duality proposal of \cite{Gaberdiel:2010pz}.\footnote{We would like to thank R. Gopakumar and S. Minwalla for an illuminating discussion on this issue.} We define the 't Hooft coupling $\lt$ as
\be
\lt (\l,c) = { \l \over \l + k(\l,c)} \label{ltitol}
\ee
with $ k(\l,c )$ given in (\ref{kl}). This  can be inverted as
\be
\l (\lt,c) = { \lt \over \lt + k(\lt,c )}.\label{litolt}
\ee
In the semiclassical limit, we follow the spectrum of primaries $(\L^+, \L^- )_\l$ when taking $c$ to be large while keeping $\l$ fixed and in the range $0\leq \l \leq1$. In this limit the energy of the primaries $(\L^+, \L^- )_\l$ behaves as in (\ref{hsc},\ref{scalperten}). Alternatively, we can consider the Gaberdiel-Gopakumar-'t Hooft limit where
we follow the primaries  $(\L^+, \L^- )_{\l ( \lt, c)}$, expressed now in terms of the 't Hooft coupling using (\ref{litolt}),  when taking $c$ to be large while keeping $\lt$ fixed and in the range $0\leq \lt \leq1$.  In this limit we find that  the energies of the primaries are (\ref{hth}) which agrees with standard expression in the   Gaberdiel-Gopakumar-'t Hooft limit of the $W_N$ minimal models (see e.g. (3.23) in \cite{Jevicki:2013kma}). All energies are positive in this limit, with the states  $(\L, \L )_{\l ( \lt, c)}$ becoming light.

Hence we see that, although the Vasiliev  theory with AdS
boundary conditions captures all states of the proposed dual CFT, it does so in classical limit which is different from the unitary 't Hooft limit. These two classical limits are in a sense strong-weak dual since we see from  (\ref{ltitol}), (\ref{litolt}) that in the semiclassical limit where $\l$ is kept fixed, the 't Hooft coupling $\lt$ becomes large and vice versa. We stress that in principle, any quantity
   computed  in the semiclassical limit can be  related to the analogous quantity computed in the 't Hooft limit and vice versa by making the substitutions (\ref{ltitol}), (\ref{litolt})  (as we illustrated for computation of the energies  in Appendix \ref{appens}), but that this requires the knowledge of the full set of  $1/c$  quantum  corrections.

Let us also comment on the importance of the triality symmetry of the $\cwl$ algebra for understanding why the two limits described above
possess the same symmetry. Indeed, the semiclassical limit is governed by $\cwl$ while the 't Hooft
limit naively seems to correspond to $\calw_\infty [ \infty ]$.
There  is however  a nontrivial `triality' isomorphism between $\calw_\infty$ algebras \cite{Gaberdiel:2012ku}, namely
\be
\cwl \simeq \calw_\infty\! \left[ {\l \over \l + k(\l,c)} \right] .
\ee
This guarantees that, for any value of $c$, the states $(\L^+, \L^- )_\l$ and $(\L^+, \L^- )_{\l /(\l + k(\l,c ))}$ are primaries of the same symmetry algebra $\cwl$. Some of the representations of $\cwl$ obtained in this way are in fact equivalent: it was argued in \cite{Gaberdiel:2011zw} that $(0, (\L^-)^T)_\l \simeq
(0, \L^-)_{\l/ (\l + k(\l,c))}$, where $\L^T$ denotes the transpose of the Young diagram $\L$, and one can check from (\ref{weightexpl}) that these have indeed the same energy. The representations $(\L^+, 0)_\l$ and
$(\L^+ , 0)_{\l/ (\l + k(\l,c))}$ however are inequivalent, since we have seen that they have different energies at large $c$. They are related
by the interpolation procedure discussed above, and it is intriguing that this exchanges perturbative  scalar quanta in one large $c$ limit
with conical solutions in the other limit.
Let us illustrate these remarks in the example of the simplest representations corresponding to Young diagrams with only one box.
There are three inequivalent such representations whose energies behave in the large $c$ limit as
\bea
h(0, \square)_\l &=&
h(0, \square)_{\l/ (\l + k(\l,c))} = \half ( 1 - \l )+ \calo ( 1/c ) \, , \\
h ( \square, 0 )_\l &=&- {c \over 2 \l^2} + \calo (1) \, , \\
h( \square,0 )_{\l/ (\l + k(\l,c))} &=& \half ( 1 + \l )+ \calo ( 1/c ) \, .
\eea
This agrees with  the direct analysis of the simplest  $\cwl$  representations from the structure of the algebra in \cite{Gaberdiel:2012ku}.

The fact that in our current setup the bulk and CFT computations give results in different regimes of the 't Hooft coupling is
 similar to what happens in other examples of holographic duality, but the fact that the large $\lt$ regime where the Vasiliev theory makes predictions is nonunitary is somewhat unsatisfying.   Therefore an interesting complementary approach to the one adopted in this work is to construct a bulk higher spin theory which directly captures the unitary 't Hooft limit of the minimal models \cite{Chang:2013izp}. This involves extending the Vasiliev theory  with new fields which are dual to those light states $(\L, \L)_{\l/(\l+k(\l,c))}$  which should be viewed as single trace operators \cite{Chang:2011vka,Jevicki:2013kma}. From the previous comments one would expect that such a theory describes, in a sense, a strong-weak 't Hooft  coupling dual of the Vasiliev theory with AdS boundary conditions.

In this work we have considered smooth solutions of the Vasiliev theory defined on a solid cylinder, where Euclidean time is noncompact. Since our solutions were time-translation and rotation invariant, we can make an extra periodic identification and consider the same solutions on  the solid torus where time runs along the non-contractible cycle.
These solutions are still smooth,  since the holonomy condition (\ref{smoothcond})  ensures that there is no singularity at the locus where the contractible $\f$-circle pinches off. They are expected to contribute to a thermal partition function (without higher spin chemical potentials turned on) in addition to the standard thermal AdS solution.
Similarly, by making the standard coordinate transformation in the bulk which reduces to a modular transformation on the boundary, we can construct
solutions on the solid torus  where the time circle is now contractible.
These `conical BTZ' solutions are smooth, the holonomy condition ensuring that there is no singularity at the locus  where the time-circle pinches off, and are again  expected to contribute to a thermal partition without higher spin chemical potentials in addition to the usual BTZ black hole. They lie on extra thermodynamic branches  which exist in higher spin gravity, discussed first for the special case $\l = 3$ in \cite{David:2012iu}, in the limit that the  chemical potentials for the higher spin fields are switched off.
 It will be interesting to see if this wealth of thermal solutions in the Vasiliev theory is a bulk manifestation of the observed absence of a Hawking-Page transition  in
the 't Hooft limit of the thermal partition function of the $W_N$ minimal models \cite{Shenker:2011zf,Banerjee:2012aj,Chen:2012ba,deBoer:2013gz}.

\section*{Acknowledgements}
We would like to thank S. Fredenhagen, R. Gopakumar, C. Iazeolla, S. Minwalla and E. Perlmutter  for useful discussions.

 The work of A.C. was partially supported by the ERC Advanced Grant ``SyDuGraM", by IISN-Belgium (convention 4.4514.08) and by the ``Communaut\'e Fran\c{c}aise de Belgique" through the ARC program. The work of J.R. has been supported  in part by the Czech Science Foundation  grant GACR P203/11/1388 and in part by the EURYI grant GACR  EYI/07/E010 from EUROHORC and ESF.  The work of T.P. has been supported  in part by the Czech Science Foundation  grant GACR P203/11/1388.

\begin{appendix}

\section{Properties of the $\hsl$ algebra}\label{apphsl}

The infinite-dimensional Lie algebra $\hsl$ is often described by exhibiting its structure constants, as we do in \eqref{star-product}.\footnote{An alternative presentation in terms of spinorial oscillator is discussed e.g.\ in \cite{Vasiliev:1989re}.} One can express the structure constants $f_\l$ appearing there in terms of those proposed in \cite{Pope:1989sr} as
\be
f_\l \! \left(
\begin{array}{ccc|cc}
i & j &&& k \\
m & n &&& m+n
\end{array}
\right) =\,
4^{-(i+j-k-1)} \, g^{\,i-1 \,,\, j-1}_{i+j-k-1} (m,n;\l) \, ,
\ee
where
\begin{align}
& g^{i\,,\,j}_k(m,n;\l) =\, \frac{1}{2(k+1)!}\, \phi^{i\,,\,j}_k(\l)\, N^{\,i\,,\,j}_k(m,n) \, ,\\[5pt]
& N^{\,i\,,\,j}_k(m,n) =\! \sum_{p\,=\,1}^{k+1}\, (-1)^p \binom{k+1}{p} (2i+2-k)_p[2j+2-p]_{k-p+1}[i+1+m]_{k-p+1}[j+1+n]_p\, , \\
& \phi^{i\,,\,j}_k(\l) = \sum_{p\,=\,0}^{\lfloor k \rfloor} \prod_{q\,=\,1}^p \frac{[(2q-3)(2q+1)-4(\l^2-1)](k-2q+3)(k/2-q+1)}{q(2i-2q+3)(2j-2q+3)(2i+2j-2k+2q+3)}\, .
\end{align}
Here $\lfloor k \rfloor$ denotes the integer part of $k$, while $(a)_n$ and $[a]_n$ denote respectively the ascending and descending Pochhammer symbols,
\begin{align}
(a)_n & \,=\, a(a+1) \ldots (a+n-1) \, , \label{poch+} \\[2pt]
[a]_n & \,=\, a(a-1) \ldots (a-n+1) \, . \label{poch-}
\end{align}

However, as we discussed in Section \ref{sec:cs}, for our goals we find it more appropriate to deal with an explicit matrix representation of $\hsl$, to be embedded in the larger $\gll$ space according to \cite{Khesin:1994ey}. More concretely, one can obtain a faithful representation of $\hsl$ working with operators acting on a base $v_i$ with $i \geq 0$ as in \eqref{gen_rep}:
\be \label{T}
T^\ell_m\, v_i = (-1)^{\ell-m} \sum_{k\,=\,0}^{\ell-m} {\ell-m \choose k} \frac{\left[\,\ell\,\right]_k}{\left[\,2\ell\,\right]_k} \left[\,\ell-\l\,\right]_k \left[\,i\,\right]_{\ell-m-k}\, v_{i+m} \, .
\ee
This expression implies $T^\ell_\ell\, v_i = (J_+)^\ell\, v_i = v_{i+\ell}$. As a result, to prove its validity one has only to verify the recursion relation \eqref{univ}. This amounts to check that $T^\ell_m$ satisfies
\be \label{[Jm,T]}
[\, J_- \,,\, T^\ell_m \,] \,=\, -\, (l + m)\, T^\ell_{m-1} \, ,
\ee
i.e.\ that it transforms as a primary state of weight $\ell$ with respect to the $sl(2)$ subalgebra. The advantages of expressing the polynomials in the ``matrix'' label $i$ in terms of Pochhammer symbols can be appreciated by noticing that any expression of the form
\be
T^\ell_m\, v_i = (-1)^{\ell-m} \sum_{k\,=\,0}^{\ell-m} {\ell-m \choose k}\, a_k(\ell,\l)\, \left[\,i\,\right]_{\ell-m-k}\, v_{i+m}
\ee
satisfies the other relevant $sl(2)$ commutation relation
\be
[\, J_+ \,,\, T^\ell_m \,] \,=\, (\ell - m)\, T^\ell_{m+1} \, .
\ee
In fact
\be
[\, J_+ \,, T^\ell_m \,]\, v_i = (-1)^{\ell-m}\! \sum_{k\,=\,0}^{\ell-m-1}\!\! {\ell-m \choose k} a_k(\ell,\l) \left\{ \left[\,i\,\right]_{\ell-m-k} - \left[\,i+1\,\right]_{\ell-m-k} \right\} v_{i+m+1}\, ,
\ee
and the identity
\be
\left[\,i\,\right]_{\ell-m-k} - \left[\,i+1\,\right]_{\ell-m-k}  = -\, (\ell-m-k) \left[\,i\,\right]_{\ell-(m+1)-k}
\ee
shifts the value of $m$ also in the binomial coefficient.
In a similar fashion one can show that \eqref{T} satisfies \eqref{[Jm,T]} because
\begin{align}
& (-1)^{\ell-m}\, [\, J_- \,, T^\ell_m \,]\, v_i \nonumber \\[2pt]
& = \sum_{k\,=\,0}^{\ell-m} {\ell-m \choose k} \frac{[\ell]_k [\ell-\l]_k}{[2\ell]_k} \Big\{ (i+m)(i-\l-m)[i]_{\ell-m-k} - i(i-\l)[i-1]_{\ell-m-k} \Big\} v_{i+m-1} \nonumber \\
& = \sum_{k\,=\,0}^{\ell-m} {\ell-m \choose k} \frac{[\ell]_k [\ell-\l]_k}{[2\ell]_k} \Big\{ (\ell+m-k)[i]_{\ell-m-k+1} + (\ell-k)(\ell-\l-k)[i]_{\ell-m-k} \Big\} v_{i+m-1} \nonumber \\
& = (\ell+m)\sum_{k\,=\,0}^{\ell-m+1} {\ell-m+1 \choose k} \frac{[\ell]_k [\ell-\l]_k}{[2\ell]_k}\, [i]_{\ell-(m-1)-k}\, v_{i+m-1} \, .
\end{align}
As an additional consistency check of our presentation of $\hsl$ one can verify that the trace of the diagonal generators $T^\ell_0$ vanishes if one computes it with the prescription \eqref{trform}:
\be
\begin{split}
\tr (T^\ell_0) & = (-1)^\ell \sum_{k\,=\,0}^\ell {\ell \choose k} \frac{[\ell]_k [\ell-\l]_k}{[2\ell]_k} \lim_{N \to \l} \sum_{i\,=\,1}^N\, [i-1]_{\ell-k} \\
& = (-1)^\ell\, [\l]_{\ell+1} \sum_{k\,=\,0}^\ell (-1)^k {\ell \choose k} \frac{\ell!(2\ell-k)!}{(\ell-k+1)!(2\ell)!} = 0 \, .
\end{split}
\ee
We also verified numerically in a number of examples that the matrix product $(T^{\,i}_m)_a{}^b (T^{\,j}_n)_b{}^c$ agrees with \eqref{star-product}, thus providing an independent check of the latter.

As we have seen, \eqref{gen_rep} or \eqref{T} are very convenient to check the agreement with the definition \eqref{univ} of the generators, but one can also present the infinite-dimensional matrices $(T^\ell_m)_{ij}$ in a more ``geometrical'' form. Such a form is given in \cite{Fradkin:1990ki} in terms of Clebsch-Gordan coefficients. In fact, one can show that \eqref{T} is equivalent to the definition given in \cite{Fradkin:1990ki} after analytic continuation from $N$ to $\lambda$ and a proper rescaling of the generators.

While moving from a more abstract presentation of $\hsl$ to an explicit matrix realization is thus straightforward for the $T^\ell_m$ generators and their finite linear combinations, we would like to stress that this is not always the case for all elements of $\gll$, as defined in \cite{Khesin:1994ey}. For instance, while one can easily express the generators $\{ T^\ell_0 \}$ in terms of the projectors $\{ P_j \}$ introduced in Section \ref{condefs}, the natural guess for inverting this relation,
\be
P_j =  {6 \over \l (\l^2 - 1)} \sum_{\ell\,=\,0}^\infty {(T^\ell_0)_{jj} \over \tr (T^\ell_0)^2}\, T^\ell_0 \, ,
\ee
is such that the would be matrix element $(P_j)_{ii}$ diverges for $\l < -\frac{3}{2}+i+j$ ($i, j = 1,2, \ldots$).

\section{Energies of the $(\L^+, \L^-)$ primaries}\label{appens}

Here we give the explicit expression for the energy of the primaries $(\L^+, \L^-)_\l$ in terms of the Young diagram data and derive the limiting expressions in the two  large $c$ limits discussed in Section \ref{disc}. The energy (\ref{weight}) of the  $(\L^+, \L^-)_\l$  primary labeled by two Young diagrams $\L^+, \L^-$ is
\bea
h(\L^+, \L^-)_\l &=& {\l + k +1 \over 2(\l + k) } {R^+ } -  {{C^+ } \over 2(\l + k) }   + {B^+ \l^2 - (B^+)^2 (\l + k + 1)\over 2 \l ( \l + k)}\nonu
&&+{\l + k  \over2( \l + k+1 )} {R^-} -  { {C^- } \over 2(\l + k+1) }  -{B^- \l^2 + (B^-)^2 (\l + k )\over 2 \l ( \l + k+1)}\nonu
&&+ {B^+ B^- \over \l } - \sum_i r^+_i r^-_i\label{weightexpl}
\eea
where as before  $r_j (c_j) $ denotes the number of boxes in the $j$-th row (column), $B$ is the total number of boxes in $\L$ and  $R = \sum_i r_i^2, \ C  = \sum_i c_i^2$. The level $k$ is the function of $\l$ and $c$ given in (\ref{kl}).

First, we discuss the semiclassical limit of this expression,  where we take $c$ to be large at fixed $\l$.
The leading part of the energy is of order $c$ and depends only on $\L^+$:
 \be
 \lim_{\tiny \begin{array}{cc}\qquad \qquad c\to \infty \\ \qquad \qquad \l fixed \end{array}} h(\L^+, \L^-)_\l = -   \left( (B^+)^2 - \l^2 B^+ + \l  (C^+ - R^+ ) \right) {c \over 2 \l^2 (1- \l^2 )} + \calo (1)\,.\label{hsc}
 \ee
 This is in agreement with the bulk calculation of the energy of the defect solutions (\ref{ensurpl}).
Another useful quantity is the energy difference between the $h(\L^+, \L^-)_\l$ and $h(\L^+, 0)_\l$ primaries:
\be
\lim_{\tiny \begin{array}{cc}\qquad \qquad c\to \infty \\ \qquad \qquad \l fixed \end{array}} \left( h(\L^+, \L^-)_\l -  h(\L^+, 0)_\l \right) = \half (C^- - B^- \l ) +   {B^+ B^- \over \l } - \sum_i r^+_i r^-_i
 + \calo (1/c)\,.\label{scalperten}
\ee
 This is positive for $|\l |\leq 1$, as can be seen from the inequalities $C\geq B, \ B^+ B^- \geq \sum_i r^+_i r^-_i$.
Hence the $h(\L^+, \L^-)_\l$ primaries can be seen as positive  energy excitations above the $h(\L^+, 0)_\l$.

Next we consider a different large  $c$ limit of (\ref{weightexpl}), namely the 't Hooft limit where instead $\lt = \l /( \l + k(\l, c))$ is kept fixed in the range $|\lt |\leq 1$. The energies (\ref{weightexpl}) in this limit are positive and of order one:
 \be
 \lim_{\tiny \begin{array}{cc}\qquad \qquad c\to \infty \\ \qquad \qquad \lt fixed \end{array}} h(\L^+, \L^-)_{\l(\l_H,\,c)} =\half    \left(  (B^+ - B^- ) |\lt | + \sum( r^+_i - r^-_i )^2 \right) + \calo (1)\,.\label{hth}
 \ee

\section{Relating highest weight and diagonal gauges}\label{gaugetransfo}
In this appendix we prove that a necessary condition for a connection in the diagonal gauge to be gauge-equivalent to a highest weight gauge connection is that all the diagonal elements are distinct and, for $\l$ non-integer, different from zero. As was shown in \cite{Campoleoni:2011hg}, the highest weight gauge is equivalent to the $u$-gauge where the connection takes the form $a_u = T^1_1 + u$, with
\be
u =
\left( \begin{array}{cccc}
u_1& u_2  & u_3 & \ldots\\
 0  &  0 & 0 &\ldots   \\
 0 &0  &0 &\dots   \\
 \vdots &         \vdots          &\vdots &\vdots
 \end{array}
 \right)\label{ugauge}
 \ee
The tracelessness condition would impose $u_1=0$, but we forget about this for the moment. The connection $a_u$ can be brought into the diagonal gauge $a_\l = \sum_j \l_j P_j$ if we can find a $g$ such that
\be
 g\,  a_\l = a_u\, g\,. \label{gtdiag}
 \ee
Here $g$ must an exponential of an element of $\hsl$. In particular, $g$ must have an inverse, a necessary condition for which is that $g$ has vanishing kernel. Writing
out (\ref{gtdiag}) as a matrix equation, one finds that it doesn't mix elements from different columns. The equations for the $j$-th column read
\bea
g_{1j} \l_j &=& \sum_{k=1}^\infty g_{kj}\, u_k \label{row1}\\
g_{kj} \l_j &=& g_{k-1\, j}\qquad k \geq 2. \label{row2}
\eea

Consider first the case $\l_j\neq 0.$ The second equation determines the column elements  in terms of $g_{1j}$:
\be
g_{kj} = { g_{1j} \over (\l_j)^{k-1}} \qquad k\geq 2. \label{solgt}
\ee
Clearly, we must have $g_{1j}$ nonzero in order for $g$ to be invertible.
Choosing the normalization $g_{1j}=1$, $g$ takes the form of an infinite Vandermonde matrix:
\be
g=
\left( \begin{array}{cccc}
1& 1  & 1 & \ldots\\
\l_1^{-1}   &  \l_2^{-1} & \l_3^{-1} &\ldots   \\
 \l_1^{-2} &\l_2^{-2}  &\l_2^{-2} &\dots   \\
 \vdots &         \vdots          &\vdots &\vdots
 \end{array}
 \right)
 \ee
 If we have two equal eigenvalues $\l_j = \l_k \neq 0$, we see that $g$ annihilates the vector $e^j -  e^k$ and hence is not invertible.
The remaining  equation (\ref{row1})
gives an infinite number of equations linking the $u_j$ and $\l_j$:
\be
\sum_{k=1}^\infty {u_k \over (\l_j)^{k} }=1.
\ee

Now let's consider the case when one of the $\l_j$ is zero. For non-integer $\l$, the second equation implies $g_{kj} = 0$ for all $k$ and $g$ again cannot be invertible. Hence also zero eigenvalues
are not allowed for generic $\l$. For integer $\l = N$, there is a caveat: if we take $g_{1j} = \l_j^N$ in (\ref{solgt}) and then take $\l_j\to 0$, the projected $gl(N)$ element is finite in the limit (however the element of the ideal diverges). Hence for $\l = N$ a single zero eigenvalue is allowed, but for two zero eigenvalues $g$ becomes again singular.

\end{appendix}

\end{document}